\documentclass[runningheads]{svmult}

\usepackage{makeidx}   
\usepackage{graphicx}  
\usepackage{subeqnar}  
\usepackage{multicol}  
\usepackage{physprbb}  
\makeindex             

\begin{document}
\title*{The Universal Equilibrium of CDM 
Halos: Making Tracks on the Cosmic Virial Plane
\footnote{to appear in "The Mass of Galaxies at Low and High Redshift"
(ESO Astrophysics Symposia), eds. R. Bender \& A. Renzini, 
Springer-Verlag, Heidelberg, in press (2002)}}
\toctitle{The Universal Equilibrium of CDM 
Halos: Making Tracks on the Cosmic Virial Plane}
\titlerunning{The Universal Equilibrium of CDM 
Halos}
%
\author{Ilian T. Iliev\inst{1}
\and Paul R. Shapiro\inst{2}}
\authorrunning{Ilian T. Iliev and Paul R. Shapiro}
%
%
\institute{Osservatorio Astrofisico di Arcetri, Largo Enrico Fermi 5, 
50125 Firenze, Italy
\and Department of Astronomy, University of Texas, Austin, 78712, USA}

\maketitle              

\begin{abstract}
Dark-matter halos are the scaffolding around which galaxies and 
clusters are built. They form when the gravitational instability of
primordial density fluctuations causes regions which are denser than average 
to slow their cosmic expansion, recollapse, and virialize. Objects as 
different in size and mass as dwarf spheroidal galaxies and 
galaxy clusters are predicted by the CDM model to have 
halos with a universal, self-similar equilibrium structure whose 
parameters
are determined by the halo's total mass and collapse 
redshift. These latter two 
are statistically correlated, however, since 
halos of the same mass form on average at the 
same epoch, with small-mass objects forming first and then 
merging hierarchically. 
 The structural properties 
of
dark-matter dominated halos of different masses, therefore, should reflect 
this statistical correlation, an imprint of the statistical properties of 
the primordial density fluctuations which formed them. Current data reveal 
these correlations, providing a fundamental test of the CDM model 
which probes the shape of the power spectrum of primordial 
density fluctuations and the cosmological background parameters. 
\end{abstract}

\section{The Truncated Isothermal Sphere (TIS) Model}
We have developed an analytical model for the postcollapse equilibrium
structure of virialized objects which condense out of a 
cosmological background universe, either matter-dominated or flat with
a cosmological constant \cite{SIR,ISb}. The model is based upon the assumption
that cosmological halos form from the collapse and virialization of
``top-hat'' density perturbations and are spherical, isotropic, and 
isothermal. This leads to a unique, nonsingular
 TIS, a particular solution of the Lane-Emden equation
(suitably modified when $\Lambda\neq0$). The size $r_t$ and velocity 
dispersion $\sigma_V$
are unique functions of the mass and redshift of formation of the object
for a given background universe. Our TIS density profile flattens to a 
constant central value, $\rho_0$, which is 
roughly proportional to the critical density of the universe at the 
epoch of collapse, with a small core radius $r_0\approx r_t/30$ (where 
$\sigma_V^2=4\pi G\rho_0r_0^2$ and 
$r_0\equiv r_{\rm King}/3$, for the ``King
radius'' $r_{\rm King}$, defined by \cite{BT}, p. 228).
The density profiles for gas and dark matter are assumed to be the same 
(no bias), with gas temperature $T=\mu m_p\sigma_V^2/k_B$.
\begin{figure}
\centering
\begin{minipage}[c]{0.45\textwidth}
\hspace{-2.2cm} 
\includegraphics[width=2.5in]{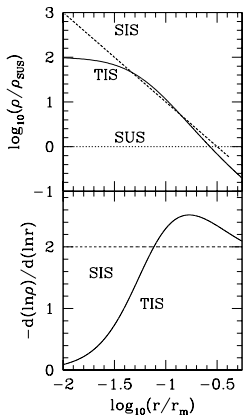}
    \label{profile}
\end{minipage}
\begin{minipage}[c]{0.4\textwidth}
\centering			
\begin{tabular}{@{}lcccc}
Table 1 &&&\\
&SUS&SIS&TIS\footnote{The two values refer to flat universe with 
$\rm \Lambda=0$ (left value) and
$\rm \Omega_0=0.3$, $\lambda_0=0.7$ (right value).}\\
\hline
$\eta/\eta_{\rm SUS}$&1&0.833&1.11;1.07\\[1mm]
$
{T}/{T_{\rm SUS}}$&1&3&2.16;2.19\\[1mm]
$\displaystyle{{\rho_0}/{\rho_t}}$& 1&$\infty$&514;530\\[1mm]
$\displaystyle{{\langle\rho\rangle}/{\rho_t}}$&1&3&3.73;3.68\\[1mm]
$\displaystyle{{r_t}/{r_0}}$& -- NA --&$\infty$&29.4;30.04\\[1mm]
${\Delta_c}/{\Delta_{\rm c,SUS}}$&1&
1.728&0.735;0.774\\[1mm]
$K/|W|$&0.5&0.75&0.683;0.690
\\\hline
\end{tabular}
\end{minipage}
\caption{(top) Density profile of TIS 
in a matter-dominated universe. Radius $r$ is in
units of $r_m$ - the top-hat radius at maximum expansion. Density $\rho$ is
in terms of the density $\rho_{SUS}$ of the SUS approximation for 
the virialized, post-collapse top-hat. (bottom) Logarithmic slope of density 
profile.} 
\end{figure}

These TIS results differ from those of the more familiar approximations in
which the virialized sphere resulting from a top-hat perturbation  
is assumed to be either the standard uniform sphere (SUS)
or else a singular isothermal sphere (SIS).
We summarize their comparison in Fig. 1 and Table 1, where $\eta$ is the 
final radius of the virialized sphere in units of the top-hat radius $r_m$
at maximum expansion (i.e. $\eta_{\rm SUS}=0.5$), $\rho_t\equiv\rho(r_t)$, 
$\langle\rho\rangle$ is the average density of the virialized spheres, 
$\Delta_c=\langle\rho\rangle/\rho_{\rm crit}(t_{\rm coll})$,
and $K/|W|$ is the ratio of total kinetic (i.e. thermal) to gravitational potential
energy of the spheres.   

\section{TIS Model vs. Numerical CDM Simulations}

The TIS model reproduces many of the average properties of the halos 
in numerical CDM simulations quite well, suggesting that it is a useful 
approximation for the halos which result from more realistic initial 
conditions: 

(1) The TIS mass profile agrees well with the fit to
N-body simulations by \cite{NFW96} (``NFW'') (i.e. fractional
deviation of $\sim20\%$ or less) at all radii outside of a few TIS core radii
(i.e. outside King radius or so), for NFW concentration parameters
$4\leq c_{\rm NFW}\leq7$ (Fig. 2). The flat density core of the TIS halo differs from
the singular cusp of the NFW profile at small radii, but this involves 
only a small fraction of the halo mass, thus not affecting their good agreement
outside the core.  
As a result, the TIS central density $\rho_0$ can be used to characterize the  
core density of cosmological halos, even if the latter have singular profiles
like that of NFW, as long as we interpret $\rho_0$, in that case,
as an average over the innermost region.
\begin{figure}
\centering
\includegraphics[height=2.2in]{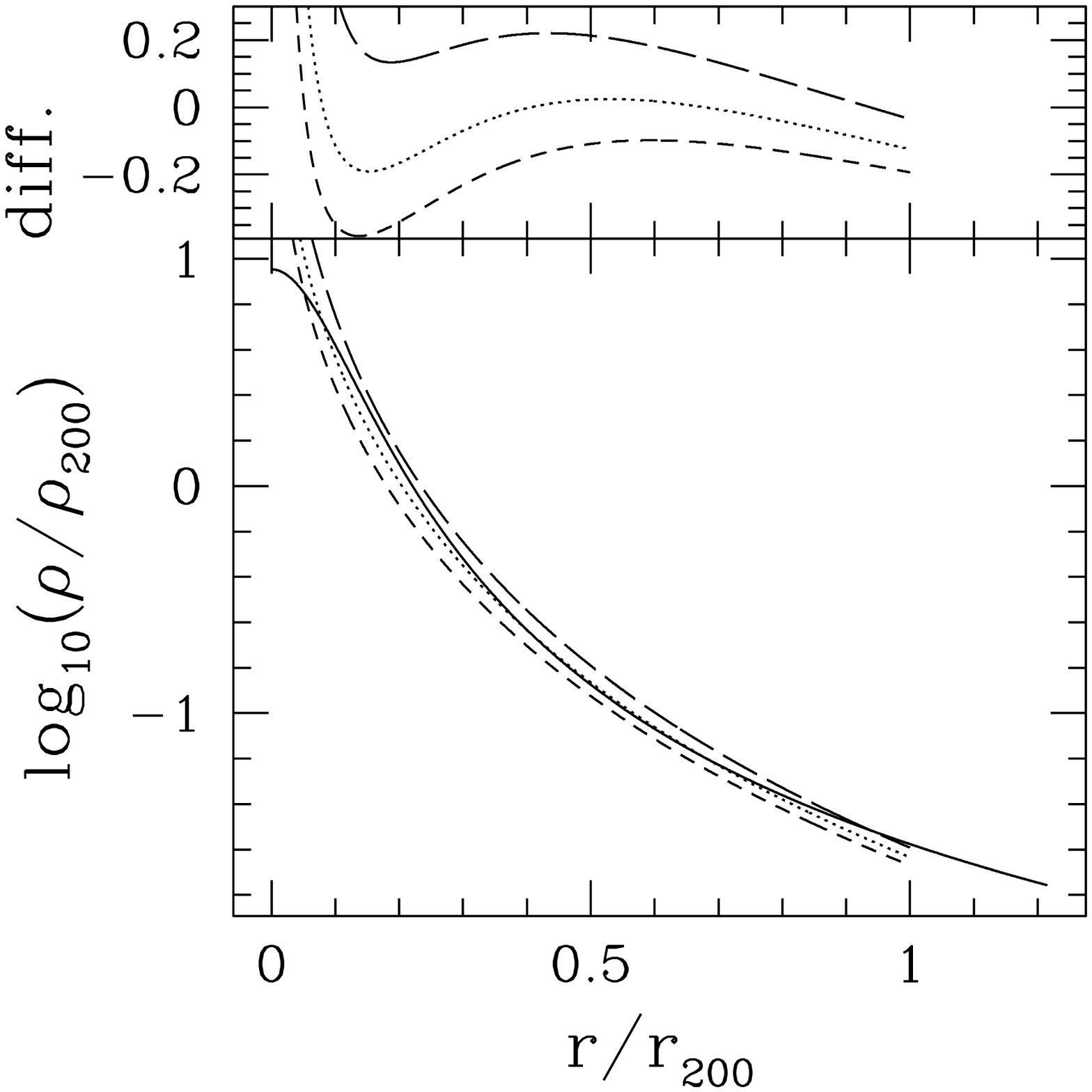}
\includegraphics[height=2.2in]{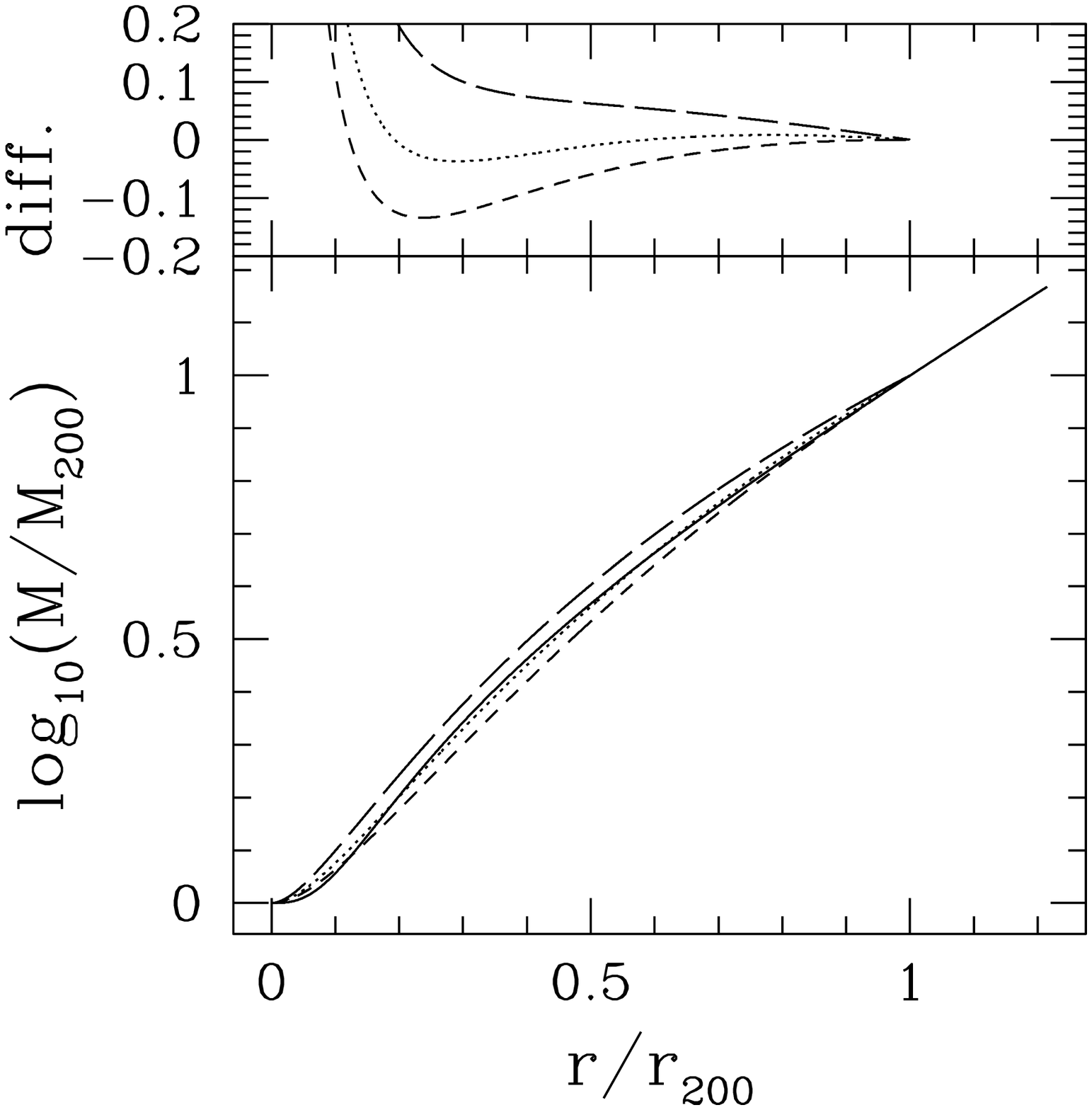}
    \label{profileNFWTIS}
\caption{Profiles of density (left) and integrated mass (right), for 
TIS (solid) and NFW with $c_{\rm NFW}=4$ (short-dashed), $5$ (dotted)
and $7$ (long-dashed) with same $(r_{200},M_{200})$.} 
\end{figure}

\indent (2) The TIS halo model predicts the internal structure of X-ray
clusters found by gas-dynamical/N-body simulations of cluster 
formation in the CDM model.  Our TIS model
predictions, for example, agree astonishingly well with the
mass-temperature and  radius-temperature virial relations and integrated 
mass profiles derived empirically from the simulations of cluster formation
by \cite{EMN,ME} (EMN).
Apparently, these simulation results are not sensitive
to the discrepancy between our
prediction of a small, finite density core and
the N-body predictions of a density cusp for 
clusters in CDM. Let $X$ be the average overdensity inside radius $r$
(in units of the cosmic mean density) $X\equiv{\langle\rho(r)\rangle}/{\rho_b}$. The
radius-temperature virial relation is defined as 
$r_X\equiv r_{10}(X)\displaystyle{\left( T/{10\, {\rm keV}}\right)^{1/2}}\,{\rm Mpc}$,
and the mass-temperature virial relation by
$M_X\equiv M_{10}(X)\displaystyle{\left( T/{10\, {\rm keV}}\right)^{1/2}}
	\,h^{-1} 10^{15}\,M_\odot$.
A comparison between our predictions of 
$r_{10}(X)$ and the results of EMN is given in Fig. 3. 
EMN obtain $M_{10}(500)=1.11\pm 0.16$ and $M_{10}(200)=1.45$, while our TIS solution
yields $M_{10}(500)=1.11$ and $M_{10}(200)=1.55$.

(3) The TIS halo model also successfully reproduces  the mass - velocity 
dispersion relation for clusters in CDM N-body simulations and 
its dependence on redshift for different background cosmologies. 
N-body simulation of the Hubble volume 
[(1000 Mpc)$^3$] by the Virgo Consortium (
reported by 
Evrard at the U. of Victoria meeting, August 2000)
yields the following empirical relation:
\begin{equation}
\sigma_V=(1080\pm65)\left[
{h(z)M_{200}}/{10^{15}M_\odot}\right]^{0.33}{\rm km/s},
\end{equation}
where $M_{200}$ is the mass within a sphere with average
density 200 times the cosmic mean density, and 
$h(z)=h_0\sqrt{\Omega_0(1+z)^3+\lambda_0}$
 is the
redshift-dependent Hubble constant. Our TIS model predicts:
\begin{equation}
\sigma_V=1103\left[
{h(z)M_{200}}/{10^{15}M_\odot}\right]^{1/3}{\rm km/s}.
\end{equation}

\begin{figure}
	\centering
    \includegraphics[width=1.9in]{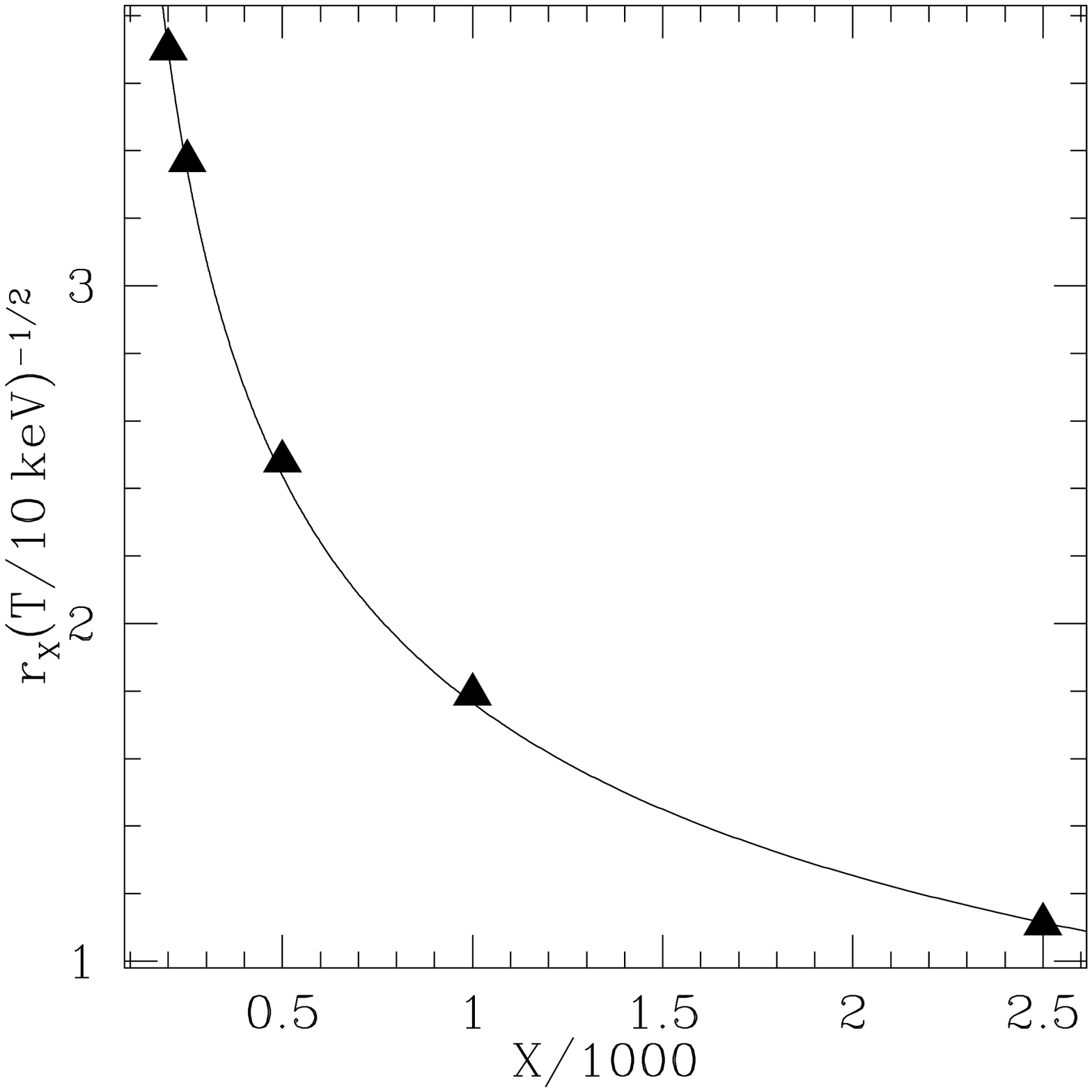}
    \includegraphics[width=1.9in]{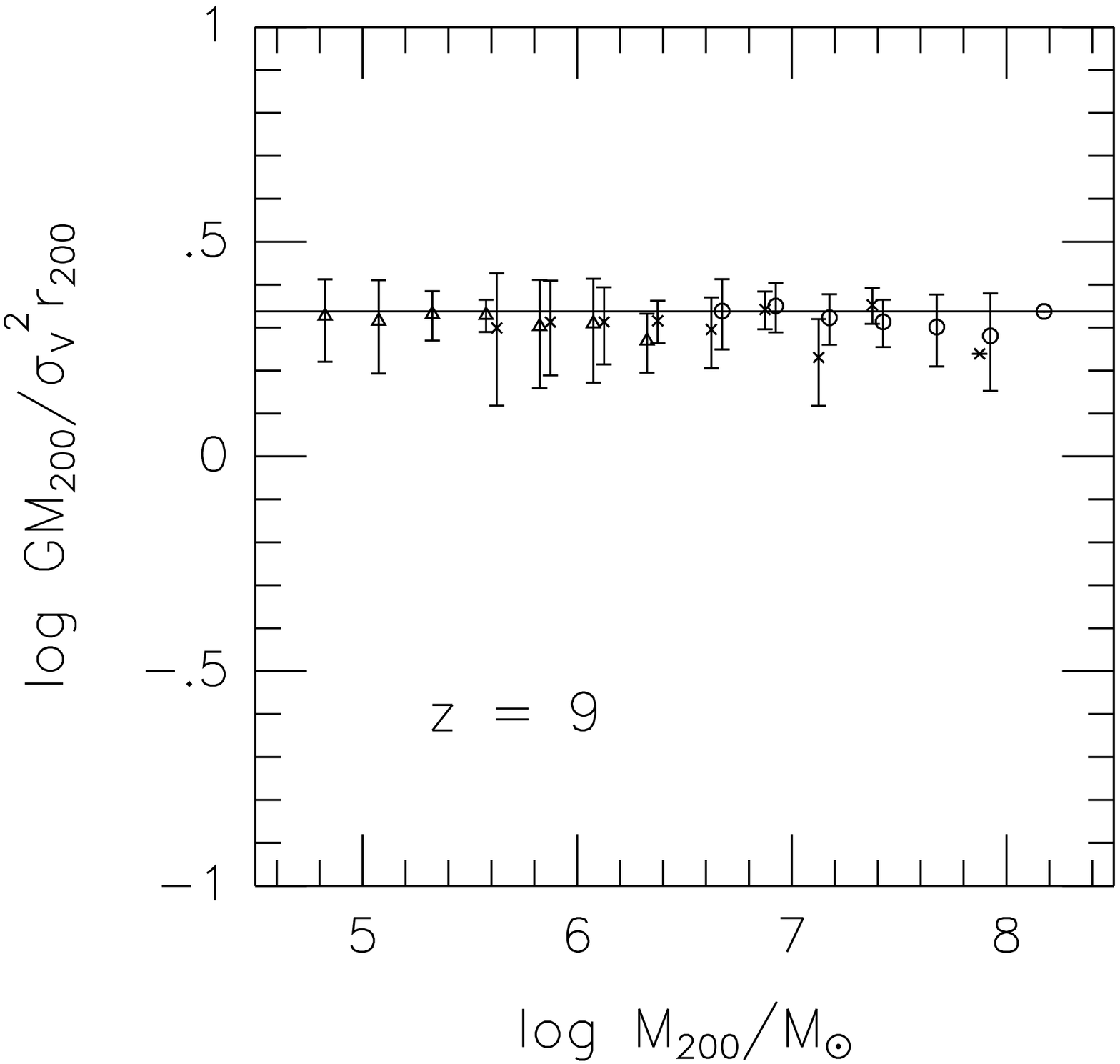}
    \caption{(a) (left) (triangles) Cluster radius-temperature virial relation
for CDM simulation results (at $z=0$) as fit by EMN and as predicted by TIS 
(solid curve).
\newline (b) $GM_{200}/(\sigma_V^2r_{200})$
vs. mass 
for halos from N-body simulations [with $1\sigma$ error bars]. 
Horizontal line is analytical prediction of TIS model.}
    \label{EMN}
\end{figure}

(4) The TIS model successfully predicts the average virial ratio,
$K/|W|$, of
halos
in CDM simulations.  An equivalent TIS
quantity, $GM_{200}/(r_{200}\sigma^2_V) = 2.176$, is 
plotted for dwarf galaxy minihalos at $z=9$ in Fig. 3(b), from \cite{S01}, 
showing good agreement between TIS and N-body halo results. A similar plot,
but of 
 $K/|W|$ for such halos, was shown by \cite{JKH01} based 
upon N-body simulations, in which the average $K/|W|$ is close to 0.7, as 
predicted by the TIS model (Table 1). Those authors were apparently unaware 
of this TIS prediction since they compared their results with the SUS value
of $K/|W|$, $0.5$, and interpreted the discrepancy incorrectly as an indication that
their halos were not in equilibrium.  

\section{TIS Model vs. Observed Halos}

(1) The TIS profile matches the observed mass profiles of 
dark-matter-dominated dwarf galaxies \cite{ISa}. 
The observed rotation curves of
dwarf galaxies are well fit by a density profile
with a finite density core given by
\begin{equation}
\rho(r)=\frac
{\rho_{0,B}}{(r/r_c+1)(r^2/r_c^2+1)}
\end{equation}
\cite{B}. The TIS model gives a nearly perfect fit to this profile (Fig.~4(a)),
with best fit parameters
$\displaystyle{{\rho_{0,B}}/{\rho_{0,TIS}}=1.216,\,
{r_{c}}/{r_{0,TIS}}=3.134}$.
This best-fit TIS profile correctly predicts $v_{\rm max}$, the maximum rotation
velocity, and the radius, $r_{\rm max}$, at which it occurs in the Burkert profile:
 $\displaystyle{{r_{\rm max,B}}/{r_{\rm max,TIS}}
	=1.13,\,}$ and  $\displaystyle{{v_{\rm max,B}}/{v_{\rm max,TIS}}=1.01}$.
\begin{figure}
	\centering
    \includegraphics[width=2.2in]{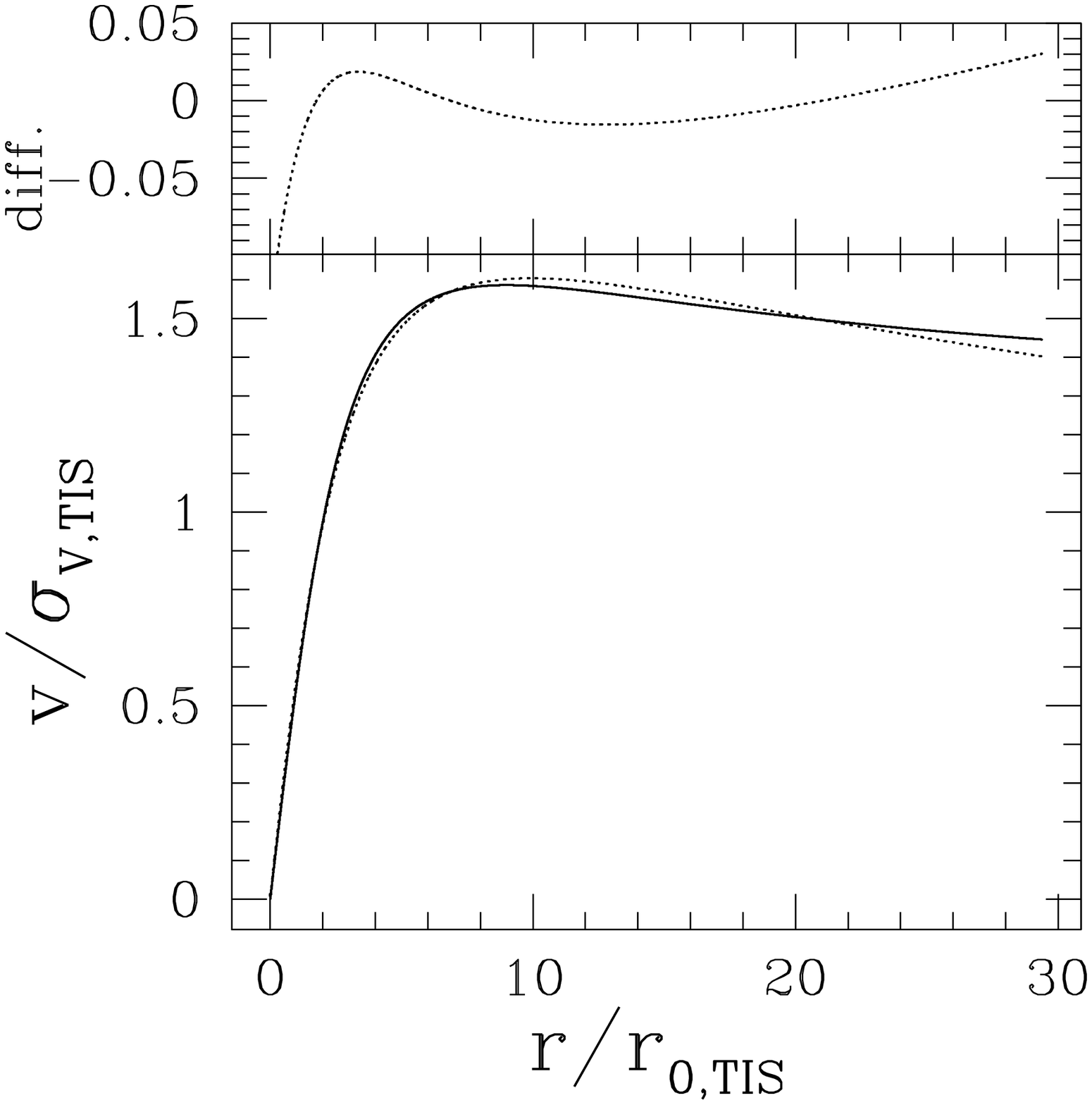}
    \includegraphics[width=2in]{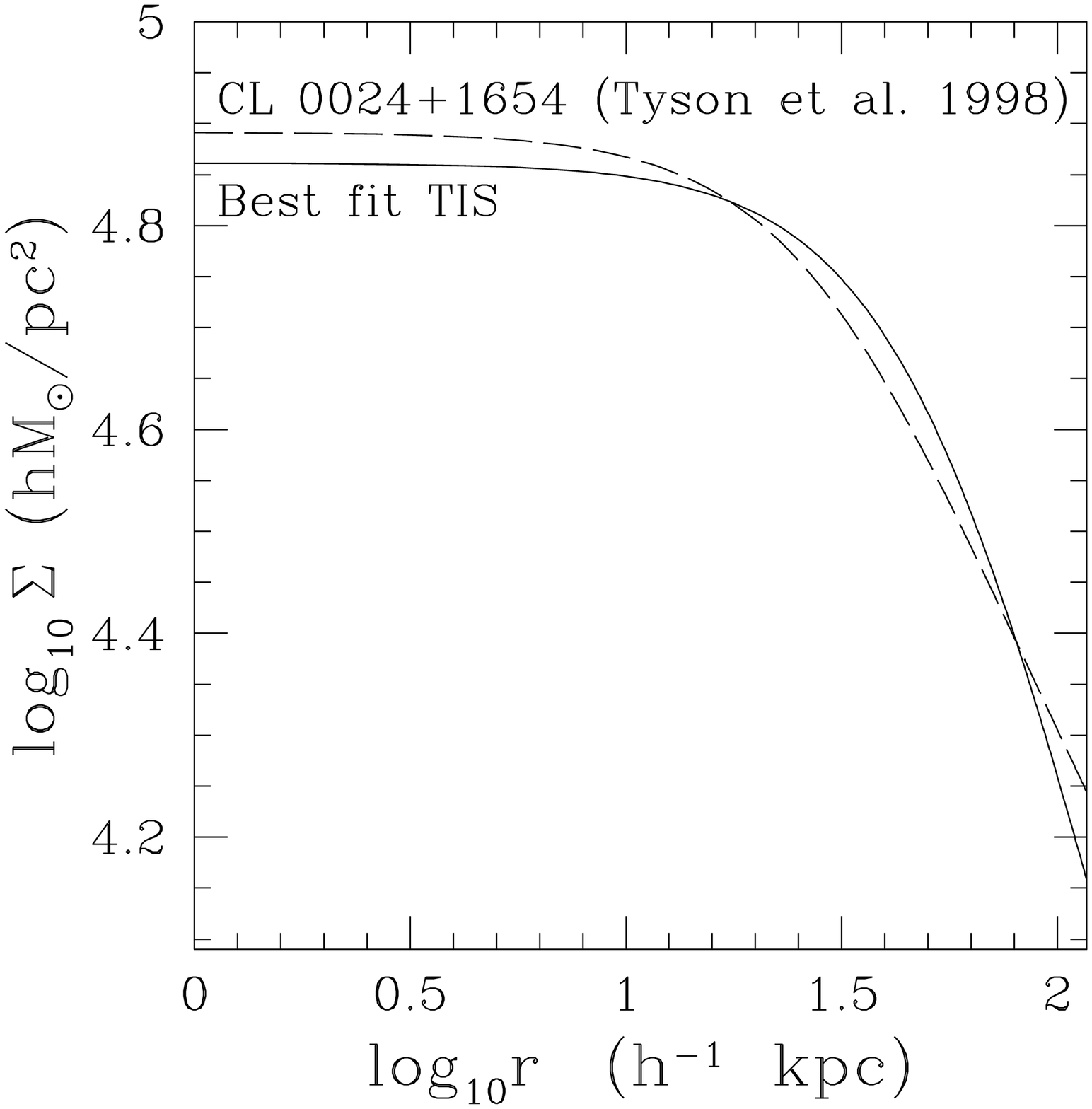}
    \caption{(a) (left) Rotation Curve Fit. 
Solid line = Best fit TIS; Dashed line = Burkert profile
(b) (right) Projected surface density of cluster CL 0024+1654 inferred from lensing 
measurements, together with the best-fit TIS model. }
    \label{burk_fit}
\end{figure}

{(2) The TIS halo model can explain the mass profile with a flat density core 
measured by \cite{TKD} for cluster CL 0024+1654
at $z=0.39$, using the strong gravitational lensing of background galaxies by the 
cluster to infer the cluster mass distribution \cite{SIa}.}
The TIS model not only provides a good fit to the projected 
surface mass density distribution of this cluster within the arcs (Fig. 4b), but
also predicts the overall 
mass, and a cluster velocity dispersion in close agreement with the value 
$\sigma_v=1150$ km/s measured by \cite{DSPBCEO}.

\section{Making Tracks on the Cosmic Virial Plane}
The TIS model yields ($\rho_0,\sigma_{\rm V},r_t,r_0$) uniquely as
functions of ($M,z_{\rm coll}$). This defines a ``cosmic virial plane" in 
($\rho_0,r_0,\sigma_{\rm V}$)-space and
determines halo size, mass, and collapse redshift for each point on the plane. 
In hierarchical clustering models like CDM, $M$ is statistically
correlated with $z_{\rm coll}$. This determines the distribution of points on 
the cosmic virial plane.
 We can combine the TIS model with the Press-Schechter (PS) approximation for
$z_{\rm coll}(M)$ -- typical collapse epoch for halo of mass $M$ -- to
predict correlations of observed halo properties.


{(1) The combined (TIS+PS) model explains the observed $v_{\rm max}$ 
-- $r_{\rm max}$ correlation of 
dwarf spiral and LSB galaxies}, with preference for the currently favoured
$\Lambda$CDM model \cite{ISa} (Fig. 5(a)).

(2) The TIS+PS model also predicts the correlations of 
central mass and phase-space densities, $\rho_0$ and 
$Q\equiv\rho_0/\sigma_V^3$, 
of dark matter halos with their velocity dispersions $\sigma_V$,
with data for low-redshift dwarf spheroidals 
to X-ray clusters again most consistent with $\Lambda$CDM 
\cite{SIb} (Fig. 5 (b,c)). 
There have been recent claims that $\rho_0=$const for all cosmological halos,
independent of their mass, as expected for certain types of SIDM
\cite{FDCHA,KKT}.
This claim, however does not seem to be supported by most current
data (Fig.~5 (c)). 
\begin{figure}
\centering
\includegraphics[height=1.65in]{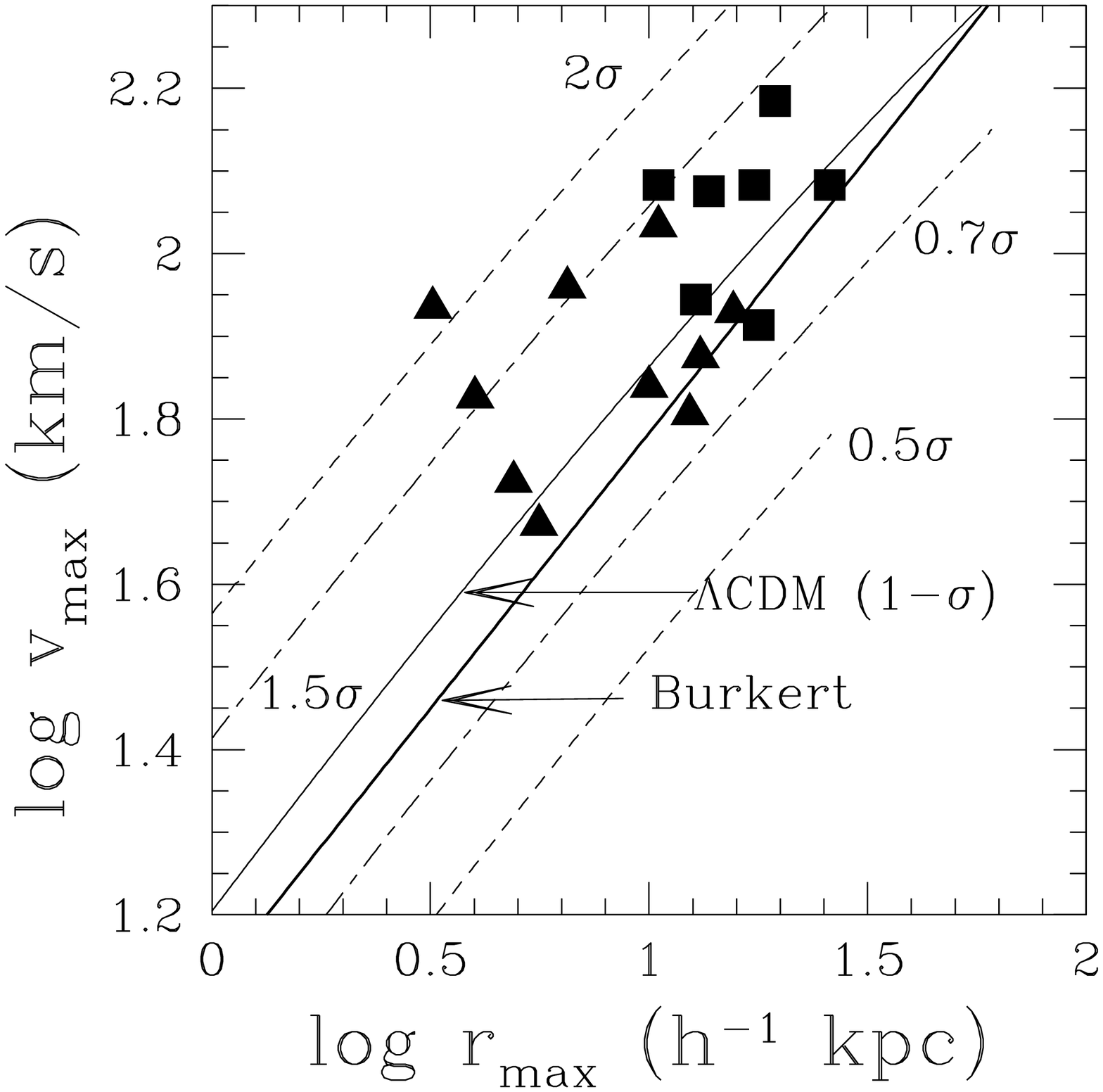}
\includegraphics[height=1.6in]{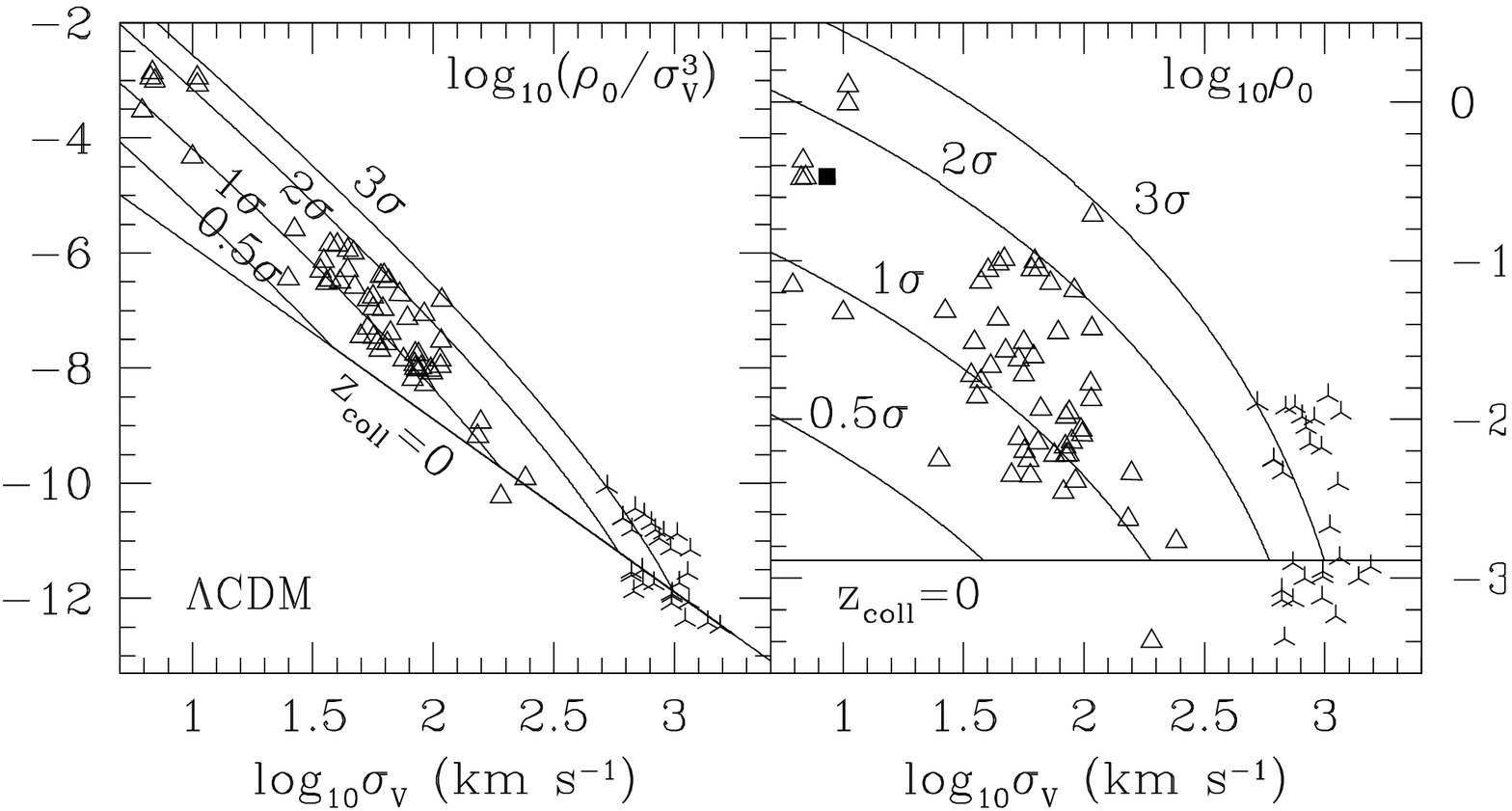}
    \label{phase_rho}
\caption{Correlations predicted by (TIS+PS) model for $\Lambda$CDM
[COBE normalized, $\Omega_0=1-\lambda_0=0.3, h=0.65$; no tilt),
for halos formed from $\nu-\sigma$ fluctuations [i.e. 
$\nu\equiv{\delta_{\rm crit}(z)}/\sigma(M,z)$, $\sigma(M,z)=$ 
standard deviation of linear density fluctuations filtered on
mass scale $M$; typical
$M(z_{\rm coll})=M(\nu=1)$], as labelled with $\nu$-values.
All curves are (TIS+PS) results, except curve in (a) labelled
``Burkert'' is a fit to data \cite{B}.
(a) (left) $v_{\rm max}$--$r_{\rm max}$ correlation.
Observed dwarf galaxies (triangles) and LSB galaxies (squares) from 
\cite{KKBP}; 
(b) (middle) $Q$--$\sigma_V$ correlation. 
 Line representing halos of different
mass which collapse at the same redshift is shown for the case 
$z_{\rm coll} = 0$.  
Data points for
galaxies and clusters are from the following: 
(1) 49 late-type spirals of type Sc-Im and 7 dSph galaxies from 
\cite{KF96,KF01} (open triangles); 
(2) Local Group dSph Leo I from \cite{MOVK} (filled square);
(3) 28 nearby clusters, $\sigma_V$ from \cite{G} and \cite{JF}, 
and $\rho_0$ from \cite{MME} (crosses).
(c) (right) Same as (b), except for 
$\rho_0$ vs. $\sigma_V$.}
\end{figure}


This work was supported by 
European Community RTN
contract HPRN-CT2000-00126 RG29185,
grants NASA NAG5-10825 and TARP 3658-0624-1999.

%

\end{document}